\documentclass[draftclsnofoot,onecolumn]{IEEEtran}
\usepackage{ifpdf}
\usepackage{cite}
\usepackage{amsmath,amssymb,amsfonts}
\usepackage{bm}
\usepackage{algorithmic}
\usepackage{array}
\usepackage{fixltx2e}
\usepackage{stfloats}
\usepackage{url}
\usepackage{graphicx}
\usepackage{textcomp}
\usepackage{wrapfig}
\usepackage{booktabs}
\usepackage{subfigure}
\usepackage{hyperref}
\usepackage{enumerate}
\usepackage{lineno}
\usepackage{multirow}
\usepackage{setspace}
\usepackage{color}
\ifCLASSINFOpdf
\else
\fi
\begin{document}
%
\title{\makebox[\linewidth]{\parbox{\dimexpr\textwidth+1.0cm\relax}{\centering A Gridless Fourth-order Cumulant-Based DOA Estimation Method under Unknown Colored Noise}}}
%
%
%

\author{Jiawen~Yuan
\thanks{This work was supported by the National Natural Science Foundation of China under Grant nos. 61871218 and 61701046, the Fundamental Research Funds for the Central University of China under Grant nos. 3082019NC2019002 and 3082017NP2017421, Funding of Postgraduate Research Practice \& Innovation Program of Jiangsu Province (KYCX20\_0201).}
\thanks{J. Yuan is with the College of Electronic and Information Engineering, Nanjing University of Aeronautics and Astronautics, Nanjing 210016, China}
}

%
%

\markboth{Journal of \LaTeX\ Class Files,~Vol.~14, No.~8, August~2015}%
{Shell \MakeLowercase{\textit{et al.}}: Bare Demo of IEEEtran.cls for IEEE Journals}
%



\maketitle

\begin{abstract}
To reduce the adverse impacts of the unknown colored noise on the performance degradation of the direction-of-arrival (DOA) estimation, we propose a new gridless DOA estimation method based on fourth-order cumulant (FOC) in this letter. We first introduce the non-redundancy single measurement vector (SMV) through FOC, which is capable of suppressing the Gaussian colored noise. Next, we analyze the distribution of the estimation error and design an estimation error tolerance scheme for it. We then combine the atomic norm minimization of the non-redundancy SMV with the above constraint scheme. This combination poses the stability of the sparsest solution. Finally, the DOA estimation is retrieved through rotational invariance techniques. Moreover, this method extends the gridless DOA estimation to the sparse linear array. Numerical simulations validate the effectiveness of the proposed method.
\end{abstract}


%
\IEEEpeerreviewmaketitle

\section{Introduction}
%
%
%
%
\IEEEPARstart{D}{irection-of-arrival} (DOA) estimation algorithms have received extensive research attention in many engineering fields such as radar $\left[1\right]$, $\left[2\right]$ and wireless communication $\left[3\right]$, $\left[4\right]$. These algorithms commonly assume that the background noise is white Gaussian. However, it is a noteworthy phenomenon that occurs in these fields, where colored noise is output as white noise after passing through linear components. This phenomenon inevitably mismatches the signal model proposed by second-order statistics-based DOA estimation methods $\left[5\right]$,$\left[6\right]$, which is the primary cause of the performance deterioration in practical applications.

Numerous efforts have been made to deal with DOA estimation in unknown colored noise $\left[7\right]$-$\left[14\right]$. Based on the assumption that the noise covariance matrix follows a certain symmetric structure $\left[7\right]$, $\left[8\right]$, it is feasible to alleviate the influence of these noises by obtaining the covariance difference matrix. However, this assumption is hard to satisfy and it fails to work in the circumstance with more sources than antennas. To overcome the above deficiencies, an algorithm based on fourth-order cumulant (FOC) $\left[9\right]$ and its variants $\left[10\right]$-$\left[14\right]$ have been proposed. All these methods have the ability to expand the effective aperture and to increase the degrees of freedom without any assumption on the structure of the noise covariance matrix. In particular, the FOC matrix-based atomic norm minimization (FOC-ANM) $\left[14\right]$ algorithm associates sparsity with the FOC matrix to offer reliable estimates under the uniform linear array (ULA) for it fundamentally eliminates the effect of basis mismatch. Compared with the grid-based DOA estimation methods $\left[4\right]$, $\left[15\right]$, FOC-ANM achieves a satisfactory estimation performance. However, FOC-ANM ignores the influence of the FOC matrix estimation error caused by the limited number of snapshots.

In this letter, we propose a new gridless FOC-based method, named error-tolerant FOC-ANM (ET-FOCANM), to enhance resolution capability and estimation accuracy under unknown colored noise. \textcolor{red}{Compared with the existing FOC-based methods, we analyze the distribution of the estimation error due to the limited number of snapshots and attach the estimation error tolerance constraint to an ANM-based model for ensuring a stable and sufficiently sparse solution. The merits of the ET-FOCANM include robustness to the Gaussian colored noise, an increased number of detectable sources in the ULA and sparse linear array (SLA) case, expanded virtual array aperture, and no requirement of user-parameter in the estimation error tolerance constraint.}

$Notations$: $\mathbb{R}$ and $\mathbb{C}$ denote the sets of numbers in real
and complex domains respectively. For a matrix $\boldsymbol{A}$, $vec$($\boldsymbol{A}$) represents the vectorization operator by taking column-wise from $\boldsymbol{A}$ and $diag$($\boldsymbol{A}$) returns a column vector of the main diagonal elements of $\boldsymbol{A}$. $\boldsymbol{A}\geq$ 0 implies that $\boldsymbol{A}$ is positive semidefinite. $\boldsymbol{A} \in \{0,1\}^{M \times N}$ means that each element of $M\times N$ dimensional matrix $\boldsymbol{A}$ is contained in the binary set $\{$0,1$\}$. For a vector $\boldsymbol{x}$, $diag$($\boldsymbol{x}$) is a diagonal matrix with $\boldsymbol{x}$ on the diagonal. $\|\cdot\|_{2}$ and $\|\cdot\|_{\mathcal{A}}$ are the $l_{2}$ and atom norms respectively. ($\cdot$)$^{T}$, ($\cdot$)$^{H}$, ($\cdot$)$^{*}$, and ($\cdot$)$^{-1}$ are the transpose, conjugate transpose, complex conjugate, and inverse operation respectively. $\otimes$ and $\odot$ denote the Kronecker and Khatri-Rao products respectively. $E\{\cdot\}$ and $cum\left(\cdot\right)$ indicate the mathematical expectation and FOC operation respectively.

\section{Signal Model and Preliminaries}
\subsection{Signal Model}
Suppose that a linear array simultaneously receives signals from $P$ narrowband far-field sources with unknown complex amplitude $s_{p}$ and distinct DOA $\theta_{p}$, $p=1,...,P$. The antenna index set of this array is defined as $\Omega=\{\Omega_{1},...,\Omega_{M}\}\subseteq\{1,...,N\}$ . For convenience, we consider two linear arrays in this letter, one is ULA with $\Omega=\{1,...,N\}$, and the other is SLA with $\Omega=\{\Omega_{1},...,\Omega_{M}\}$ of $\Omega_{1}=1$ and $\Omega_{M}=N \left[5\right], \left[6\right]$. Fig. 1 shows a configuration of the 4-element ULA with $\Omega=\{1,2,3,4\}$, 4-element SLA with $\Omega=\{1,2,5,7\}$, and 7-element ULA with $\Omega=\{1,...,7\}$, where the inter-antenna spacing is taken as equaling half of the wavelength.
\begin{figure}[!t]
\centering
\includegraphics[height=1.27cm,width=5.77cm]{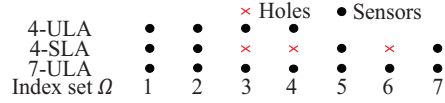}
\caption{An example of the 7-element ULA, 4-element ULA and SLA.}
\label{fig1}
\end{figure}

\textcolor{red}{The received signals $\boldsymbol{Y}_{\Omega}=\left[\boldsymbol{y}_{\Omega}(1),...,\boldsymbol{y}_{\Omega}(J)\right]\in\mathbb{C}^{M \times J}$ with multiple snapshots $J$ in the $M$-element SLA case can be expressed as follows
\begin{equation}
\boldsymbol{Y}_{\Omega}=\boldsymbol{\Gamma}_{\Omega} \boldsymbol{Y}=\boldsymbol{A}_{\Omega} \boldsymbol{S}+\boldsymbol{N}_{\Omega} \quad with \quad \boldsymbol{Y}=\boldsymbol{A} \boldsymbol{S}+\boldsymbol{N},
\end{equation}
where $\boldsymbol{\Gamma}_{\Omega}\in\!\!\{0,1\}^{M \times N}$,  $\boldsymbol{A}_{\Omega}=\left[\boldsymbol{a}_{\Omega}(\theta_{1}),..., \boldsymbol{a}_{\Omega}(\theta_{P})\right] \in \mathbb{C}^{M \times P}$, $\!\!\boldsymbol{S}=\left[\boldsymbol{s}(1),..., \boldsymbol{s}(J)\right] \in \mathbb{C}^{P \times J}$, and $\boldsymbol{N}_{\Omega}=\left[\boldsymbol{n}_{\Omega}(\theta_{1}),..., \boldsymbol{n}_{\Omega}(\theta_{J})\right] \in \mathbb{C}^{M \times J}$ denote the selection matrix, the array manifold matrix, the source signal, and the additive noise, respectively.} Let $\boldsymbol{a}_{\Omega}(\theta_{p})=\left[e^{j\pi(\Omega_{1}-1)\sin\theta_{p}},...,e^{j\pi(\Omega_{M}-1)\sin\theta_{p}}\right]^{T}$ be the $p$-th steering vector. In the $N$-element ULA, the definitions of $\boldsymbol{Y}, \boldsymbol{N} \in \mathbb{C}^{N \times J}$, and $\boldsymbol{A} \in \mathbb{C}^{N \times P}$ are similar to the previous ones. Moreover, we assume that the source is uncorrelated stationary non-Gaussian with zero-mean, and the noise is the zero-mean colored Gaussian independent of the sources.

\subsection{Definition of reduce-complexity FOC (RC-FOC) matrix}
Based on the above assumptions, the received signals are considered as stationary non-Gaussian processes with zero mean. Since the FOC is blind to Gaussian processes $\left[16\right]$, we construct the FOC matrix $\boldsymbol{C}_{4} \in \mathbb{C}^{N^{2} \times N^{2}}$ of $\boldsymbol{y}(t)\in \mathbb{C}^{N}$ as
\begin{IEEEeqnarray}{Rl}
\boldsymbol{C}_{4}&=E\left\{\left(\boldsymbol{Y}\otimes\boldsymbol{Y}^{*}\right)\left(\boldsymbol{Y}\otimes\boldsymbol{Y}^{*}\right)^{H}\right\}-E\left\{\boldsymbol{Y}\otimes\boldsymbol{Y}^{*}\right\} \IEEEnonumber\\
&\quad\times E\left\{\boldsymbol{Y}\otimes\boldsymbol{Y}^{*}\right\}^{H}-E\left\{\boldsymbol{Y} \boldsymbol{Y}^{H}\right\}\otimes E\left\{\boldsymbol{Y} \boldsymbol{Y}^{H}\right\}^{*}.
\end{IEEEeqnarray}
Then $\boldsymbol{C}_{4}$ can be decomposed as
\begin{IEEEeqnarray}{Rl}
\boldsymbol{C}_{4}=\left(\boldsymbol{A} \odot \boldsymbol{A}^{*} \right)\boldsymbol{C}_{\boldsymbol{S}} \left(\boldsymbol{A} \odot \boldsymbol{A}^{*}\right)^{H} = \overline{\boldsymbol{B}}\boldsymbol{C}_{\boldsymbol{S}} \overline{\boldsymbol{B}}^{H},
\end{IEEEeqnarray}
where $\boldsymbol{C}_{\boldsymbol{S}}=diag(\boldsymbol{\gamma})\in\mathbb{C}^{P\times P}$, $\boldsymbol{\gamma}=\left[\gamma_{4 s_{p}},...,\gamma_{4 s_{P}}\right]^{T} \in \mathbb{C}^{P}$,  and its $p$-th element $\gamma_{4s_{p}}=cum\left(s_{p},s_{p}^{*}, s_{p}^{*}, s_{p}\right)$.

Due to a large number of redundant elements in $\boldsymbol{C}_{4}$, the RC-FOC matrix $\boldsymbol{R}_{4}\in\mathbb{C}^{(2N-1)\times (2N-1)}$ is designed in $\left[14\right]$
\begin{IEEEeqnarray}{Rl}
\boldsymbol{R}_{4}&=\overline{\boldsymbol{G}}^{-1}\overline{\boldsymbol{H}}^{T}\boldsymbol{C}_{4}\overline{\boldsymbol{H}} \overline{\boldsymbol{G}}^{-1}\IEEEnonumber\\
&=\overline{\boldsymbol{G}}^{-1} \overline{\boldsymbol{H}}^{T}\overline{\boldsymbol{B}} \boldsymbol{C}_{\boldsymbol{S}}\overline{\boldsymbol{B}}^{H} \overline{\boldsymbol{H}} \overline{\boldsymbol{G}}^{-1}=\boldsymbol{B} \boldsymbol{C}_{\boldsymbol{S}} \boldsymbol{B}^{H},
\end{IEEEeqnarray}
where $\overline{\boldsymbol{G}}=\overline{\boldsymbol{H}}^{T} \overline{\boldsymbol{H}}=diag \left\{1,..., N,N-1,...,1\right\}$ and $\overline{\boldsymbol{H}}^{T}=\left[\overline{\boldsymbol{H}}_{1}^{T}, \overline{\boldsymbol{H}}_{2}^{T}, ..., \overline{\boldsymbol{H}}_{N}^{T}\right] \in \mathbb{R}^{(2N-1) \times N^{2}}$ is an orthogonal matrix. Moreover, the definition of $\overline{\boldsymbol{H}}_{n}$ is represented as
\begin{equation}
\!\overline{\boldsymbol{H}}_{n}=\left\{\begin{array}{cc}{\left[\mathbf{0}_{N \times(N-1)}, \boldsymbol{I}_{N}\right]} & \!n=1 \\ {\left[\mathbf{0}_{N \times(N-n)}, \boldsymbol{I}_{N}, \mathbf{0}_{N \times(n-1)}\right]} & \!\quad 2 \!\leq\! n \leq \!\!N\!\!-\!\!1\!\! \\ {\left[\boldsymbol{I}_{N}, \mathbf{0}_{N \times(N-1)}\right]} & \!n=N.\end{array}\right.
\end{equation}
$\boldsymbol{B}=\left[\boldsymbol{b}(\theta_{1}),...,\boldsymbol{b}(\theta_{P})\right] \in \mathbb{C}^{(2N-1) \times P}$ is the new array manifold matrix with the $p$-th column vector $\boldsymbol{b}(\theta_{p})=\left[e^{-j\pi(N-1)\sin\theta_{p}},...,e^{j\pi(N-1)\sin\theta_{p}}\right]^{T}$.

\section{The ET-FOCANM Algorithm}
\subsection{Introduction of non-redundancy SMV}
Consider a ULA composed of $N$ omnidirectional antennas, since rank($\boldsymbol{R}_{4}$) is tied to the number of sources in the ULA case, the condition $P < (4N-3)$ can be taken as a priori information. Given $\boldsymbol{R}_{4}$ as in $\left(4\right)$, let us define the vector $\overline{\boldsymbol{z}}$.
\begin{IEEEeqnarray}{Rl}
\overline{\boldsymbol{z}} &=vec\left(\boldsymbol{R}_{4}\right)=\left(\boldsymbol{B}^{*} \odot \boldsymbol{B}\right) \boldsymbol{\gamma}\IEEEnonumber\\
&=\sum\nolimits_{p = 1}^P \gamma_{4 s_{p}} \boldsymbol{b}\left(\theta_{p}\right)^{*} \odot \boldsymbol{b}\left(\theta_{p}\right)=\boldsymbol{HD\gamma},
\end{IEEEeqnarray}
where $\boldsymbol{D}=\left[\boldsymbol{d}(\theta_{1}),...,\boldsymbol{d}(\theta_{P})\right]\in \mathbb{C}^{(4N-3) \times P}$, and $\boldsymbol{d}(\theta_{p})=\left[e^{-j\pi(2N-2)\sin\theta_{p}},...,e^{j\pi(2N-2)\sin\theta_{p}}\right]^{T}$. It can be seen that $\overline{\boldsymbol{z}}$ still has the repeating entries. Therefore, we choose a new orthogonal matrix $\boldsymbol{H} \in \mathbb{C}^{(2N-1)^{2} \times (4N-3)}$ whose form is consistent with $\overline{\boldsymbol{H}}$ to perform the linear transformation on $\overline{\boldsymbol{z}}$
\begin{equation}
\boldsymbol{z}=\boldsymbol{G}^{-1} \boldsymbol{H}^{T} \overline{\boldsymbol{z}}=\boldsymbol{D\gamma}=\sum\nolimits_{p = 1}^P \gamma_{4 s_{p}} \boldsymbol{d}\left(\theta_{p}\right),
\end{equation}
\noindent where $\boldsymbol{z} \in \mathbb{C}^{4N-3}$ is the non-redundancy single measurement vector (SMV) and $\boldsymbol{G} = \boldsymbol{H}^{T}\boldsymbol{H}$. As a byproduct, it can generate $O\left\{4N-3\right\}$ degrees of freedom from only $O\left\{N\right\}$ antennas, which is beneficial to detect more sources.
\subsection{Estimation error tolerance scheme}
Since finite snapshots inevitably lead to estimation error in the actual situation, it has the potential to relax the above equality constraint $\left(7\right)$ and demand instead
\begin{equation}
\boldsymbol{z}=\boldsymbol{D}\boldsymbol{\gamma}+\boldsymbol{\varepsilon}=\boldsymbol{x}+\boldsymbol{\varepsilon},
\end{equation}
in which $\boldsymbol{\varepsilon}$ is the estimation error vector from non-redundancy SMV. As shown in $\left[5\right]$, the sparse model $\boldsymbol{x}$ can be linearly represented by $P$ atoms in a set of the continuous atoms, so we make the following optimization to reconstruct $\boldsymbol{x}$
\begin{equation}
\hat{\boldsymbol{x}}=\underset{\boldsymbol{x}}{\arg \min }\|\boldsymbol{x}\|_{\mathcal{A}} \quad s.t.\quad \|\boldsymbol{\varepsilon}\|_{2}^{2}=\|\boldsymbol{z}-\boldsymbol{x}\|_{2}^{2} \leq \xi,
\end{equation}
\noindent where $\xi$ denotes the tolerance of error energy. Unlike the constraint in $\left[14\right]$, this inequality is controlled by $\xi$, where $\xi$ is a result of multiple factors, including signal-to-noise ratio (SNR) and snapshot. Thus, it is hard to select a proper value of $\xi$ for the final DOAs.

To proceed, we innovatively propose a new estimation error tolerance scheme that does not depend on $\xi$ but on the statistical property. First, let $\Delta \boldsymbol{C}_{4}=\hat{\boldsymbol{C}_{4}}-\boldsymbol{C}_{4}$ be the estimation error matrix of FOC and $\hat{\boldsymbol{C}_{4}}$ denote the sample observation of $\boldsymbol{C}_{4}$. \textcolor{red}{Inspired by $\left[17\right]$, we deduce the distribution of error component $vec(\Delta\boldsymbol{C}_{4})$ in the complex domain, which is expressed by
\begin{equation}
vec\left(\Delta \boldsymbol{C}_{4}\right) \sim AsN(\boldsymbol{0}, \boldsymbol{V}).
\end{equation}
$AsN(\boldsymbol{0}, \boldsymbol{V})$ denotes asymptotic normal distribution with zero mean and the covariance matrix $\boldsymbol{V}$.} Meanwhile, the element of $\boldsymbol{V}$ can be determined by
\begin{equation}
\label{eqn_dbl_y}
\begin{array}{l}
\!\!\!\!cov\left\{\hat{c}_{4}(\tau),\hat{c}_{4}(\rho)\right\}\!\!=\!\!\frac{1}{J}\!\!\left\{Q_{44}(\tau ; \rho)
-\sum_{v=1}^{3}\!\!\left[\!\!\begin{array}{l}
Q_{42}\left(\tau ; k_{1}^{\prime}, k_{v 1}^{\prime}\right)\!\!E\{k_{v 2}^{\prime}k_{v 3}^{\prime}\}\!+\!Q_{42}\left(\tau ; k_{v 2}^{\prime}, k_{v 3}^{\prime}\right)\!\! E\{k_{1}^{\prime}k_{v 1}^{\prime}\} \\
+Q_{42}^{*}\left(\rho ; k_{1}^{\prime}, k_{v 1}^{\prime}\right)\!\!E\{k_{v 2}k_{v 3}\}^{*}\!+\!Q_{42}^{*}\left(\rho ; k_{v 2}^{\prime}, k_{v 3}^{\prime}\right)\!\!E\{k_{1}k_{v 1}\}^{\!*}
\end{array}\!\!\right]\right. \\
\!\!+\!\!\left.\sum_{\mu,\!v=\!1}^{3}\!\!\!\left[\!\!\!\!\begin{array}{l}
Q_{22}\left(k_{1}, k_{v1} ; k_{1}^{\prime}, k_{\mu 1}^{\prime}\right)\!\!E\{k_{v2}k_{v3}\}^{*}\!\!E\{k_{\mu2}^{\prime}k_{\mu 3}^{\prime}\}^{\!*}\!\!+\!\!Q_{22}\left(k_{v2}, k_{v3} ; k_{1}^{\prime}, k_{\mu1}^{\prime}\right)\!\!E\{k_{1} k_{v1}\}^{\!*}\!\!E\{k_{\mu2}^{\prime}k_{\mu3}^{\prime}\}^{\!*} \\
\!+\!Q_{22}\left(k_{1}, k_{v1} ; k_{\mu2}^{\prime}, k_{\mu3}^{\prime}\right)\!\!E\{k_{v2}k_{v3}\}^{\!*}\!\!E\{k_{1}^{\prime}k_{\mu1}^{\prime}\}^{\!*}\!\!+\!\!Q_{22}\left(k_{v2}, k_{v3} ; k_{\mu 2}^{\prime}, k_{\mu3}^{\prime}\right)\!\!E\{k_{1}k_{v 1}\}^{\!*}E\{k_{1}^{\prime}k_{\mu 1}^{\prime}\}^{\!*}
\end{array}\!\!\right]\!\!\right\}\!\!.
\end{array}
\end{equation}
where $\tau\!=\!\left\{k_{1},\!k_{2},\!k_{3},\!k_{4}\right\}$, $\rho\!\!=\!\!\left\{\!k_{1}^{'},\!k_{2}^{'},\!k_{3}^{'},\!k_{4}^{'}\!\right\}$, $k_{11}\!=\!k_{2}$, $k_{12}\!=\!k_{3}$, $k_{13}\!=\!k_{4}$, $k_{21}\!=\!k_{3}$, $k_{22}\!=\!k_{2}$, $k_{23}\!=\!k_{4}$, $k_{31}\!=\!k_{4}$, $k_{32}\!=\!k_{3}$, $k_{33}\!=\!k_{2}$ and the same as $k^{'}$. \textcolor{red}{$E\{k_{1}k_{v1}\}$ is the expectation of $y_{k_{1}}y_{k_{v1}}$ and the same as $E\{k_{1}^{'}k_{v1}^{'}\}$, $E\{k_{v2}k_{v3}\}$, $E\{\!k_{v2}^{'}k_{v3}^{'}\!\}$, $E\{\!k_{1}^{'}k_{\mu1}^{'}\!\}$, $E\{\!k_{\mu2}^{'}k_{\mu3}^{'}\!\}$. The definition of $Q_{ij}$ is as follows
\begin{IEEEeqnarray}{Rl}
\!\!Q_{ij} \!=\! \mathop {\lim }\limits_{J \to \infty} \!\! Jcov \{\hat{m}_{i},\hat{m}_{j}\}\!=\!\!\!\!\sum\limits_{\xi=-\infty }^{\infty}\!\!\!cov \{{f_k}(t),f_{{k^{'}}}(t + \xi)\},
\end{IEEEeqnarray}
where ${f_k}(t) = {y_{{k_1}}}(t)y_{{k_2}}^{*}(t) \cdots {y_{{k_i}}}(t)$ and ${f_{{k^{'}}}}(t) = {y_{k_1^{'}}}(t)y_{k_2^{'}}^{*}(t) \cdots {y_{k_j^{'}}}(t)$, $i,j \le 4$. $y_{{k_i}}(t)$ denotes the $k_i$-th element of $\boldsymbol{y}(t)$ with $k_i \in \left[1,N\right]$, and the same as $y_{k_j^{'}}(t)$.}

Next, utilizing the orthogonal invariance property of Gaussian random matrix, it can be inferred as
\begin{IEEEeqnarray}{Rl}
\boldsymbol{\varepsilon}=& \boldsymbol{G}^{-1} \boldsymbol{H}^{T} vec\left(\Delta \boldsymbol{R}_{4}\right)=\boldsymbol{G}^{-1} \boldsymbol{H}^{T} vec\left(\widehat{\boldsymbol{R}}_{4}-\boldsymbol{R}_{4}\right) \IEEEnonumber\\
=& \boldsymbol{G}^{-1} \boldsymbol{H}^{T}\left(\overline{\boldsymbol{G}}^{-1} \overline{\boldsymbol{H}}^{T}\right) \otimes\left(\overline{\boldsymbol{G}}^{-1} \overline{\boldsymbol{H}}^{T}\right) vec\left(\Delta \boldsymbol{C}_{4}\right) \IEEEnonumber\\
=& \boldsymbol{W} vec\left(\Delta \boldsymbol{C}_{4}\right) \sim AsN\left(0, \boldsymbol{W} \boldsymbol{V} \boldsymbol{W}^{H}\right).
\end{IEEEeqnarray}
\noindent It directly results in
\begin{equation}
\left\|\boldsymbol{W}^{-\frac{1}{2}} \boldsymbol{\varepsilon}\right\|_{2}^{2} \sim As\chi^{2}(4N-3),
\end{equation}
\noindent where $As\chi^{2}(4N-3)$ denotes the asymptotic chi-square distribution with $4N-3$ degrees of freedom.

Finally, based on the property of $\chi^{2}$ distribution, the following inequality holds with a high probability $1-\delta$ where $\delta=0.001$ is enough:
\begin{equation}
\left\|\boldsymbol{W}^{-\frac{1}{2}} \boldsymbol{\varepsilon}\right\|_{2}^{2} \leq \eta.
\end{equation}
\noindent The choice of $\eta$ is easily calculated by the code $chi2inv(1-\delta, 4N-3)$ in Matlab.
\subsection{ANM-based model attached to the above constraint}
We propose the error-tolerant problem by substituting the (15) into the constraint term of (9)
\begin{equation}
\hat{\boldsymbol{x}}=\underset{\boldsymbol{x}}{\arg \min }\|\boldsymbol{x}\|_{\mathcal{A}} \quad s.t.\quad \left\|\boldsymbol{W}^{-\frac{1}{2}} \boldsymbol{\varepsilon}\right\|_{2}^{2} \leq \eta.
\end{equation}
\noindent This problem must always yield the optimal solution at least as sparse as that in (9). \textcolor{red}{Further, the result of (16) can be solved by the SDP problem $\left[18\right]$
\begin{IEEEeqnarray}{c}
\{\hat{\boldsymbol{x}}, \hat{q}, \boldsymbol{T}(\hat{\boldsymbol{\mu}})\}=\underset{\boldsymbol{x}, q, T(\boldsymbol{\mu})}{\arg \min } \frac{1}{2} q+\frac{1}{2} \mu_{1} \IEEEnonumber\\
s.t. \left\|\boldsymbol{W}^{-\frac{1}{2}} \boldsymbol{\varepsilon}\right\|_{2}^{2} \leq \eta,\left[\begin{array}{cc}\boldsymbol{T}(\boldsymbol{\mu}) & \quad \boldsymbol{x} \\ \boldsymbol{x}^{H} & \quad q\end{array}\right] \geq 0,
\end{IEEEeqnarray}
where $q$ denotes the smallest dilation factor and $\boldsymbol{T}\left(\boldsymbol{\mu}\right)$ represents the Hermitian Toeplitz matrix with its first column $\boldsymbol{\mu}=\left[\mu_{1},...,\mu_{4N-3}\right]^{T}$ and rank($\boldsymbol{T}\left(\boldsymbol{\mu}\right))=P \le (4N-4) $. Hence, when the convex optimization toolbox called CVX $\left[19\right]$ produces the optimal solution $\boldsymbol{T}\left(\boldsymbol{\mu}\right)$ of (17), it is easy to estimate the DOAs through some general techniques such as Vandermonde decomposition $\left[5\right]$, rotational invariance techniques (ESPRIT) $\left[20\right]$, spectral peak search $\left[9\right]$ and its variant root-MUSIC $\left[21\right]$. To simplify the calculation, we perform rotational invariance techniques on the ensuing DOA estimation.}

\section{The Extension to an SLA}
In this subsection, we extend ET-FOCANM method to the SLA case. For simplicity, we mainly focus on one type of the SLA, which is the minimum redundancy array (MRA) $\left[5\right]$, $\left[6\right]$. Therefore, according to the signal model (1) and the definition of $\boldsymbol{C}_{4}$, the FOC matrix $\boldsymbol{C}_{4\Omega} \in \mathbb{C}^{M^{2} \times M^{2}}$ is designed as follows
\begin{IEEEeqnarray}{Rl}
\!\!\!\!\!\!\!\!\!\!\!\!\boldsymbol{C}_{4 \Omega}&=E\left\{\left(\boldsymbol{Y}_{\Omega} \otimes \boldsymbol{Y}_{\Omega}^{*}\right)\left(\boldsymbol{Y}_{\Omega} \otimes \boldsymbol{Y}_{\Omega}^{*}\right)^{H}\right\}\!-\!E\left\{\boldsymbol{Y}_{\Omega} \otimes \boldsymbol{Y}_{\Omega}^{*}\right\} \IEEEnonumber\\
&\!\quad\!\times\! E\left\{\boldsymbol{Y}_{\Omega} \!\otimes\! \boldsymbol{Y}_{\Omega}^{*}\right\}^{H}\!-\!E\left\{\boldsymbol{Y}_{\Omega} \boldsymbol{Y}_{\Omega}^{H}\right\} \otimes E\left\{\boldsymbol{Y}_{\Omega} \boldsymbol{Y}_{\Omega}^{H}\right\}^{*}.
\end{IEEEeqnarray}
\noindent Then we derive $\boldsymbol{R}_{4}$ under the SLA case
\begin{equation}
\!\!\!\boldsymbol{R}_{4}=\boldsymbol{G}_{\Omega}^{-1}\left[\left(\boldsymbol{\Gamma}_{\Omega} \otimes \boldsymbol{\Gamma}_{\Omega}\right) \overline{\boldsymbol{H}}\right]^{T} \!\boldsymbol{C}_{4 \Omega}\!\left[\left(\boldsymbol{\Gamma}_{\Omega} \otimes \boldsymbol{\Gamma}_{\Omega}\right) \overline{\boldsymbol{H}}\right] \boldsymbol{G}_{\Omega}^{-1},
\end{equation}
where $\boldsymbol{G}_{\Omega}=\left[\left(\boldsymbol{\Gamma}_{\Omega} \otimes \boldsymbol{\Gamma}_{\Omega}\right) \overline{\boldsymbol{H}}\right]^{T}\left[\left(\boldsymbol{\Gamma}_{\Omega} \otimes \boldsymbol{\Gamma}_{\Omega}\right) \overline{\boldsymbol{H}}\right]$. The acquisition of $\boldsymbol{z}$ is the same as in (8) and let $\Delta \boldsymbol{C}_{4\Omega}=\hat{\boldsymbol{C}}_{4\Omega}-\boldsymbol{C}_{4\Omega}$ be the FOC matrix error in the SLA case. \textcolor{red}{Since $vec(\boldsymbol{C}_{4\Omega})$ is equal to $\left[\left(\boldsymbol{\Gamma}_{\Omega} \otimes \boldsymbol{\Gamma}_{\Omega}\right)\otimes \left(\boldsymbol{\Gamma}_{\Omega} \otimes \boldsymbol{\Gamma}_{\Omega}\right)\right]vec(\boldsymbol{C}_{4})$,  which reveals that $vec(\boldsymbol{C}_{4\Omega})\sim AsN(\boldsymbol{0}, \boldsymbol{V}_{\Omega})$. Further, we conclude the link between $\boldsymbol{\varepsilon}$ and $\Delta \boldsymbol{C}_{4\Omega}$
\begin{IEEEeqnarray}{c}
\!\!\boldsymbol{\varepsilon}=\boldsymbol{W}_{\Omega}vec\left(\Delta \boldsymbol{C}_{4 \Omega}\right) \sim AsN\left(\boldsymbol{0}, \boldsymbol{W}_{\Omega} \boldsymbol{V}_{\Omega} \boldsymbol{W}_{\Omega}^{H}\right),
\end{IEEEeqnarray}
where $\boldsymbol{W}_{\Omega}\!\!\!\!=\!\!\!\!\boldsymbol{G}^{-1}\boldsymbol{\!H}^{T}(\boldsymbol{G}_{\Omega}^{-1}\left(\left(\boldsymbol{\Gamma}_{\Omega}\otimes\boldsymbol{\Gamma}_{\Omega}\right)\overline{\boldsymbol{H}}\right)^{T}\otimes\boldsymbol{G}_{\Omega}^{-1}\left(\left(\boldsymbol{\Gamma}_{\Omega}\otimes\boldsymbol{\Gamma}_{\Omega}\right)\overline{\boldsymbol{H}}\right)^{T})$. }Similarly, we have
\begin{equation}
\left\|\boldsymbol{W}_{\Omega}^{-\frac{1}{2}} \boldsymbol{\varepsilon}\right\|_{2}^{2} \sim A s \chi^{2}(4N-3).
\end{equation}
\noindent The denoising problem in the SLA case is rewritten as follows:
\begin{equation}
\widehat{\boldsymbol{x}}=\underset{\boldsymbol{x}}{\arg \min }\|\boldsymbol{x}\|_{\mathcal{A}} \quad s.t. \left\|\boldsymbol{W}_{\Omega}^{-\frac{1}{2}} \boldsymbol{\varepsilon}\right\|_{2}^{2} \leq \eta,
\end{equation}
\noindent and the problem (22) is equivalent to
\begin{IEEEeqnarray}{c}
\{\hat{\boldsymbol{x}}, \hat{q}, \boldsymbol{T}(\hat{\boldsymbol{\mu}})\}=\underset{\boldsymbol{x}, q, T(\boldsymbol{\mu})}{\arg \min } \frac{1}{2} q+\frac{1}{2} \mu_{1} \IEEEnonumber\\
s.t. \left\|\boldsymbol{W}_{\Omega}^{-\frac{1}{2}} \boldsymbol{\varepsilon}\right\|_{2}^{2} \leq \eta,\left[\begin{array}{cc}\boldsymbol{T}(\boldsymbol{\mu}) & \quad \boldsymbol{x} \\ \boldsymbol{x}^{H} & \quad q\end{array}\right] \geq 0.
\end{IEEEeqnarray}
To get an accurate estimate of $\boldsymbol{T}(\boldsymbol{\mu})$ through CVX, we use rotational invariance techniques again to retrieve the DOAs.

\section{Simulation Results}
In this subsection, compared with MUSIC-LIKE $\left[9\right]$, JDS-FOC $\left[11\right]$ and FOC-ANM $\left[14\right]$ algorithms, we provide several simulations to demonstrate the effectiveness of the proposed method. Consider the ULA and redundancy SLA composed of $4$ omnidirectional antennas in Fig. 1. Furthermore, the non-Gaussian sources are modeled as $\boldsymbol{s}\left(t\right)=\boldsymbol{F}\left(t\right)\boldsymbol{e}\left(t\right)$ where $\boldsymbol{F}(t)=diag\{f_1(t),...,f_P(t)\}$ and $\boldsymbol{e}(t)=\left[e_1(t),...,e_P(t)\right]^{T}$. The zero-mean Gaussian processes $f_p(t)$ and $e_p(t)$ are with unit-variance and $\sigma_{s}^{2}$-variance  $\left[13\right]$, respectively. The noise is generated by Gaussian white noise through a second-order autoregressive filter $\left[22\right]$ with the coefficients $\left[1,-1,0.8\right]$. For each simulation, $100$ Monte Carlo trials are collected.
\begin{figure}[!h]
\begin{spacing}{0.01}
\centering
\includegraphics[height=4.77cm,width=5.57cm]{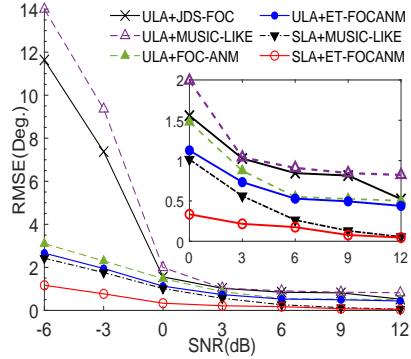}
\caption{RMSE of DOA estimation versus SNR with $J=300$.}
\label{fig1}
\end{spacing}
\end{figure}
\begin{figure}[!h]
\begin{spacing}{0.01}
\centering
\includegraphics[height=4.77cm,width=5.59cm]{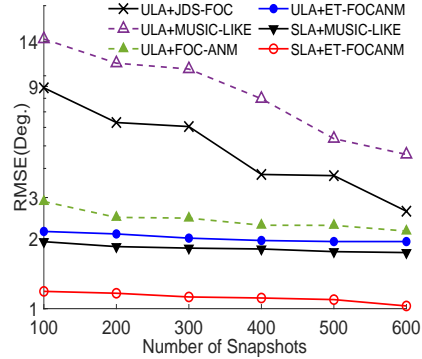}
\caption{RMSE of DOA estimation versus snapshots with SNR = $-3$ dB.}
\label{fig1}
\end{spacing}
\end{figure}

To illustrate the estimation precision performance of ET-FOCANM, we employ the root mean square error (RMSE) as a metric to evaluate estimation precision. The DOAs of the uncorrelated signals are $\theta_{1}=-23^{\circ}$ and $\theta_{2}=17^{\circ}$ in the first simulation. Fig. 2 plots the RMSE curve as SNR varies from $-6$ dB to $12$ dB with $3$ dB as the step. Fig. 3 depicts the RMSE curve as snapshots $J$ varies from $100$ to $600$ with $100$ as the step. It is indicated in Fig. 2 and Fig. 3 that ET-FOCANM outperforms the other methods in the ULA case due to its appropriate error energy constraints. Moreover, the resulting RMSEs verify that the estimation error of ET-FOCANM applied to SLAs is much smaller than that applied to ULAs under the same antenna scale.

\section{Conclusion}
In this letter, we propose a gridless FOC-based method, named ET-FOCANM, for DOA estimation in the two linear arrays. After suppressing the colored noise by FOC, ET-FOCANM implements the practicable estimation error tolerance scheme and is converted into an ANM-based form. The experimental results indicate that ET-FOCANM can achieve better angular precision and resolution capability than the traditional algorithms. For future work, it will be worth applying the alternating direction method of multipliers to speed up the solution of the ANM-based form in ET-FOCANM.

\ifCLASSOPTIONcaptionsoff
  \newpage
\fi




\begin{thebibliography}{1}
\bibitem{1}
Y.~Fang, S.~Zhu and Y.~Gao, “Direction Finding by Covariance Matrix Sparse Representation With Sensor Gain and Phase Uncertainties in Unknown Non-Uniform Noise,” \emph{IEEE Wireless Commun. Lett.}, vol. 10, no. 1, pp. 175-179, Jan. 2021.
\bibitem{2}
Y.~Zhang, G.~Zhang and H.~Leung, “Gridless coherent DOA estimation based on fourth-order cumulants with Gaussian colored noise,” \emph{IET Radar Sonar Navig.}, vol. 14, no. 5, pp. 677-685, Feb. 2020.
\bibitem{3}
H.~Chen, W.~Wang and W.~Liu, “Joint DOA, Range, and Polarization Estimation for Rectilinear Sources With a COLD Array,” \emph{IEEE Wireless Commun. Lett.}, vol. 8, no. 5, pp. 1398-1401, Oct. 2019.
\bibitem{4}
Y.~Tian and H.~Xu, “Calibration nested arrays for underdetermined DOA estimation using fourth-order cumulant,” \emph{IEEE Commun. Lett.}, vol. 24, no. 9, pp. 1949-1952, Sept. 2020.
\bibitem{5}
Z.~Yang and L.~Xie, “On gridless sparse methods for line spectral estimation from complete and incomplete data,” \emph{IEEE Trans. Signal Process.}, vol. 63, no. 12, pp. 3139–3153, Jul. 2015.
\bibitem{6}
X.~Wu, W.~Zhu and J.~Yan, “A Toeplitz covariance matrix reconstruction approach for direction-of-arrival estimation,” \emph{IEEE Trans. Veh. Technol.}, vol. 66, no. 9, pp. 8223-8237, Apr. 2017.
\bibitem{7}
S.~Prasad et al., “A transform-based covariance differencing approach for some classes of parameter estimation problems,” \emph{IEEE Trans. Acoust. Speech Signal Process.}, vol. 36, no. 5, pp. 631–641, May 1988.
\bibitem{8}
F.~Wen, J.~Shi, and Z.~Zhang, “Direction finding for bistatic MIMO radar with unknown spatially colored noise,” \emph{Circuits Syst. Signal Process.}, vol. 39, no. 4, Feb. 2020.
\bibitem{9}
B.~Porat and B.~Friedlander, “Direction finding algorithms based on high-order statistics,” \emph{IEEE Trans. Signal Process.}, vol. 39, no. 9, pp. 2016–2024, Sep. 1991.
\bibitem{10}
C.~Liu, Z.~Ye, and Y.~Zhang, “DOA estimation based on fourth-order cumulants with unknown mutual coupling,” \emph{Signal Process.}, vol. 89, no. 9, pp. 1839-1843, Sep. 2009.
\bibitem{11}
W.~Zeng, X.~Li, and X.~Zhang, “Direction-of-arrival estimation based on the joint diagonalization structure of multiple fourth-order cumulant matrices,” \emph{IEEE Signal Process. Lett.}, vol. 16, no. 3, pp. 164-167, Mar. 2009.
\bibitem{12}
B.~Liao and S.~Chan, “A cumulant-based method for direction finding in uniform linear arrays with mutual coupling,” \emph{IEEE Antennas Wirel. Propag. Lett.}, vol. 13, pp. 1717-1720, Nov. 2014.
\bibitem{13}
J.~Liu, W.~Zhou, and X.~Wang, “Fourth-order cumulants-based sparse representation approach for DOA estimation in MIMO radar with unknown mutual coupling,” \emph{Signal Process.}, vol. 128, pp. 123–130, Nov. 2016.
\bibitem{14}
Y.~Zhang, G.~Zhang and H.~Leung, “Gridless sparse methods based on fourth-order cumulant for DOA estimation,” in \emph{Proc. IGARSS}, Yokohama, Japan, 2019, pp. 3416-3419.
\bibitem{15}
B.~Wang, J.~Liu and X.~Sun, “Mixed sources localization based on sparse signal reconstruction,” \emph{IEEE Signal Process. Lett.}, vol. 19, no. 8, pp. 487-490, Aug. 2012.
\bibitem{16}
N.~Yuen and B.~Friedlander, “DOA estimation in multipath: an approach using fourth-order cumulants,” \emph{IEEE Trans. Signal Process.}, vol. 45, no. 5, pp. 1253-1263, May 1997.
\bibitem{17}
A.~V. Dandawate and G.~B. Giannakis, “Asymptotic properties and covariance expressions of kth-order sample moments and cumulants,” in \emph{Proc. ACSSC}, Pacific Grove, CA, USA, 1993, pp. 1186-1190.
\bibitem{18}
G.~Tang et al., “Compressed sensing off the grid,” \emph{IEEE Trans. Inf. Theory}, vol. 59, no. 11, pp. 7465-7490, Nov. 2013.
\bibitem{19}
M.~Grant and S.~Boyd, “CVX: Matlab software for disciplined convex programming, version 2.0 beta,” \emph{Available online at http://cvxr.com/cvx}, Sep. 2013.
\bibitem{20}
J.~Lin et al., “Time-Frequency Multi-Invariance ESPRIT for DOA Estimation,” \emph{IEEE Antennas Wirel. Propag. Lett.}, vol. 15, pp. 770-773, Mar. 2016.
\textcolor{red}{\bibitem{21}
B.~D.~Rao and K.~V.~S.~Hari, “Performance analysis of Root-Music,” \emph{IEEE Trans. Acoust., Speech, Signal Proces.}, vol. 37, no. 12, pp. 1939-1949, Dec. 1989.}
\bibitem{22}
H.~Jiang, J.~Zhang, and K.~Wong, “Joint DOD and DOA estimation for bistatic MIMO radar in unknown correlated noise,” \emph{IEEE Trans. Veh. Technol.}, vol. 64, no. 11, pp. 5113–5125, Nov. 2015.
\end{thebibliography}
%

\end{document}